\documentclass[9pt]{article}
\usepackage{newinutile}
\usepackage{soul}
\usepackage{amsmath,amssymb,amsthm,amsfonts}
\usepackage{dsfont}
\usepackage{epsfig,graphics,color,calc,graphicx}
\usepackage{subfigure}
\graphicspath{{Figures/}}
\usepackage{comment}
\usepackage{pgf,tikz}
\usepackage{mathrsfs}
\usepackage{color}
\usetikzlibrary{arrows}
\usepackage{cancel}
\usepackage{mathrsfs}
\usepackage{hyperref}
\hypersetup{colorlinks=true,linkcolor=blue}
\usepackage{placeins}
\usepackage{mathtools}

\newcommand{\cc}[1]{{\mathcal{#1}}}

\newcommand\be{\begin{equation}}
\newcommand\ee{\end{equation}}
\newcommand\bea{\begin{eqnarray}}
\newcommand\eea{\end{eqnarray}}

\newcommand\zl{\lambda}

\newcommand\ensavg{\left\langle e^{-\beta W_d} \right\rangle}

\def\lsim{\,\lower2truept\hbox{${< \atop\hbox{\raise4truept\hbox{$\sim$}}}$}\,}
\def\gsim{\,\lower2truept\hbox{${> \atop\hbox{\raise4truept\hbox{$\sim$}}}$}\,}

\newlength{\pecettawidth}
\newcommand{\pecetta}[1]{
	\medskip
	\begin{equation}gin{center}
		\framebox{
			\begin{array}gin{minipage}{\pecettawidth}\texttt{Commento.}\ #1\end{minipage}
		}
	\end{center}
	\medskip
}
\begin{document}

\title{Detecting Phase Transitions through Nonequilibrium Work Fluctuations}

\author{Matteo Colangeli}
	\email{matteo.colangeli1@univaq.it}
	\affiliation{Dipartimento di Ingegneria e Scienze dell'Informazione e
		Matematica,
		Universit\`a degli Studi dell'Aquila, via Vetoio, 67100 L'Aquila, Italy.}
	
	\author{Antonio Di Francesco}
	\email{antonio.difrancesco5@graduate.univaq.it}
	\affiliation{Dipartimento di Ingegneria e Scienze dell'Informazione e
		Matematica,
		Universit\`a degli Studi dell'Aquila, via Vetoio, 67100 L'Aquila, Italy.}

	\author{Lamberto Rondoni}
	\email{lamberto.rondoni@polito.it}
	\affiliation{Dipartimento di Scienze Matematiche,
		Politecnico di Torino, Corso Duca degli Abruzzi 24, 10129 Torino, Italy \\
  INFN, Sezione di Torino, Via Pietro Giuria 1, 10125 Torino, Italy \\
  ORCID: 0000-0002-4223-6279 }

\date{\today}
	
\begin{abstract}
We show how averages of exponential functions of path dependent quantities, such as those of Work Fluctuation Theorems, detect phase transitions in
deterministic and
stochastic systems. 
State space truncation --the restriction of the observations to a subset of state space with prescribed probability-- is introduced to obtain that result. Two stochastic 
processes undergoing first-order phase transitions are analyzed both analytically and numerically: the Ehrenfest urn model and the 2D Ising model subject to a magnetic field. 
In presence of phase transitions, we prove that even minimal
state space truncation
makes averages of exponentials of path dependent variables sensibly
deviate from full state space values. Specifically, in the case of discontinuous phase transitions, this approach is strikingly effective in locating the critical transition value of the control parameter.
As this approach works even with 
variables different from those of fluctuation theorems, it provides a new recipe to identify order parameters in the study of nonequilibrium phase transitions, profiting from the often incomplete statistics that are available.

\end{abstract}
	
\maketitle

\section{Introduction}
\label{s:Intro}
\par\noindent

Exponentials of microscopically expressed variables have been used for a long time in statistical physics and molecular dynamics. Bennett's formulae for the free energy \cite{bennett1976efficient},
Widom's relation \cite{widom1963some}, and
Zwanzig's relation \cite{zwanzig1955high}, 
are among them.
More recently, they have become popular in nonequilibrium statistical mechanics in the form of Fluctuation Relations \cite{ECM2,jepps2016dynamical},
which constitute a step in the direction of generalizing linear response theory and the Fluctuation Dissipation Relation,
and in the form of Work Fluctuation Theorems (WFT) \cite{jarzynski1997nonequilibrium,crooks1998nonequilibrium,crooks2000,hatano2001steady}, that focus on equilibrium properties derived from nonequilibrium processes. See Refs.\ \cite{Bettolo,Seifert_2012}
for comprehensive reviews.

Exponential observables, such as $e^{-\beta\sigma}$ with $\sigma$ denoting a kind of dissipation functional of the system trajectories, may  provide useful insight in the properties of nonequilibrium systems, but
need to be handled with care. To the very least, 
the amount of data for accurate calculations may be prohibitively large, because substantial contributions to the  statistical properties of such observables pertain to the (low probability) tails of the relevant distributions, 
cf.\ {\em e.g.}\
Ref.\cite{jarzynski1997nonequilibrium}. The problem is often enhanced by high energy barriers, that keep trajectories from accessing specific regions. The problem is well known. For instance, in the context of free energy differences estimation \cite{bennett1976efficient,hansen2014practical}, 
numerous \textit{ad hoc} techniques and procedures have been developed to deal with it \cite{torrie1977nonphysical,pohorille2010good,wu2005phase}. In general, many challenges  remain.
Moreover, the recent work 
\cite{colangeli2023finite},
concerning the canonical ensemble of classical Hamiltonian systems, shows how exponential variables  can be strongly affected by 
finite size effects or lack of ergodicity \cite{colangeli2023finite}.
This must be considered, in particular, when dealing with small systems, because ensembles were originally conceived to treat specific observables of large systems, that: a) are not affected by
finite size effects; and b) their time averages equal their phase space averages. 

In this paper, we will show how the above difficulties can actually be leveraged to recover important information about the system behavior. This way, often unavoidable hindrances in the investigation of the objects of modern interest may turn profitable, under some conditions.

We start concisely recalling the theoretical background
of WFT for Markov processes \cite{hack2022jarzynski}, and  the notion of absolute irreversibility
\cite{murashita2014nonequilibrium}.
Then we introduce the concepts of state space truncation, that can be interpreted both as a
result of incomplete information on the system of interest, or as the representation of a system with finitely many degrees of freedom, interacting with only a portion of its environment. 
This is of interest, for instance, when the system is subjected to rapid transformations, which is common in present day bio- and nano-technologies.
We find that the corresponding ensemble
averages 
of $e^{-\beta W_d}$,
$W_d$ being the dissipated work, may detect phase transitions and locate them in parameters space.

This is shown analyzing in detail
a variant of the celebrated Ehrenfest urn model, derived from a deterministic 2D model of particles transport \cite{cirillo2020deterministic},
and 
a classical 2D Ising model with periodic boundary conditions in an external magnetic field.
For both it is found that the ensemble average of $e^{-\beta W_d}$ as a function of the driving control parameter, computed over 
trajectories starting in the lowest energy states (thus, with highest stationary probability), manifests a discontinuity in the thermodynamic limit whenever the system undergoes a first order phase transition. A smooth dependence is instead found in the case of 
the Ising model at supercritical temperatures. 

The paper is structured as follows: in Sec.\ \ref{s:TheoryAndMethods} we introduce the theoretical background and the ensemble averages over relevant subsets of the state space, along with the notion of absolute irreversibility. Section \ref{s:models} analyzes in detail the systems undergoing  phase transitions mentioned above, using a suitably amended version of the Jarzynski Equality.
Conclusions are drawn in Sec.\ \ref{s:Conclusions}.

\section{Theory and Methods}
\label{s:TheoryAndMethods}
\par\noindent

\subsection{Work and free energy for Markov Chains}
\label{WFEMC}
Here, we concisely recall basic concepts referring to discrete time Markov chains.
Let $\tau>0$ be an integer number. Denote by
$P\left(X_i=x_i|X_0=x_0,...,X_{i-1}=x_{i-1}\right)$ the probability that $x_i$ follows 
as the $i$-th step of a trajectory that started at $x_0$, went through $x_1$ at the first step, and eventually reached $x_{i-1}$ at the $(i-1)$-th step. The corresponding transition matrix $Q=(Q_{xy})$ has entries $Q_{xy} = P(X_i=y|X_{i-1}=x)$, for all $x,y\in S$ and $1\le i \le \tau$, representing the transition probabilities from the state $x$ to the state $y$, which obey 
$ \sum_{y\in S}Q_{xy}=1$, for all $x\in S$. 
A Markov Chain is a finite set of random variables $\mathbf{X}=(X_i)_{i=0}^{\tau}$, taking values on a finite state space $S$, such that the following
$$
P\left(X_i=x_i|X_0=x_0,...,X_{i-1}=x_{i-1}\right)= P\left(X_i=x_i|X_{i-1}=x_{i-1}\right) \; ,
$$
holds for all $0\leq i \leq \tau$, and for all
$(x_0,x_1,...,x_{\tau})\in S^{\tau+1}$.
The chain $\mathbf{X}$ can be characterised by the probability of the path
$\mathcal{P}_{\mu_0}(\mathbf{x})=\mu_0(x_0)\mathcal{P}(\mathbf{x}|x_0)= \mu_0(x_0)\prod_{i=0}^{\tau-1}Q_{x_ix_{i+1}}$, where
$\mu_0(x)=P(x_0=x)$ is an assigned discrete probability distribution. 
A Markov chain admits a unique invariant distribution $\mu$ on $S$, such that $\mu Q=\mu$ (where $\mu$ is conceived as a row vector) if it is irreducible and its states are ergodic, {\em i.e.}\ they are aperiodic and persistent with finite mean recurrence time \cite{Feller}. 

An explicit time dependence is sometimes introduced in the definition of the transition matrices to mimic the effect of an external drive affecting the evolution of the system. The action of the drive is expressed by a time dependent parameter $\lambda=\lambda(t)$, that follows a given \textit{protocol} $\lambda(i)=\lambda_i$, $i=0,1,...,\tau$. This way, the Markov chain turns inhomogeneous, namely
$(Q_i)_{xy}=P(X_i=y|X_{i-1}=x)$ depends on time. 

One may then introduce the following path dependent quantity called \textit{work}
\cite{Sekimoto}:
$$
W=\sum_{i=0}^{\tau-1}\left[ E_{i+1}(x_i)-E_{i}(x_i)\right] 
$$
where the function $E_i(x_i)$ is related to a distribution $\mu_0(x)$ supported on $S$ via:
\begin{equation}
   \mu_0(x)=\frac{1}{Z_i}e^{-\beta E_i(x)}\; , \qquad x\in S \;,
   \label{eq:E_def}
\end{equation}
with $Z_i=\sum_{x\in S} \exp{\{-\beta E_i(x)\}}$, for some $\beta>0$. Correspondingly, the quantity $F=-\beta^{-1}\log Z$ is called 
\textit{free energy} of the system. Therefore, letting $\Delta F \equiv F_{\tau}-F_0=-\beta^{-1}\log Z_{\tau}/Z_0$ be the free energy difference between the states $x_{\tau}$ and $x_0$ \cite{hack2022jarzynski} and introducing $W_d=W-\Delta F$ as the \textit{dissipated work}, 
the Jarzynski Equality (JE) \cite{jarzynski1997nonequilibrium} attains the form:
\begin{equation}
\left\langle e^{-\beta W_d(\mathbf{x})} \right\rangle = 1 \; ,
\label{eq:JarzEq_toprove}
\end{equation}
where, for a generic path-dependent observable $\mathcal{A}:S^{\tau+1}\rightarrow\mathbb{R}$, the angular brackets represent the following average with respect to the initial measure:
$$\langle \mathcal{A}(\mathbf{x})\rangle =\sum_{\mathbf{x}\in S^{\tau+1}} \mathcal{A}(\mathbf{x})\mu_0(x_0) \mathcal{P}(\mathbf{x}|x_0) \; .
$$ 

For the quantity of interest in Eq.\eqref{eq:JarzEq_toprove}, the following can be established:
\begin{eqnarray}
    e^{-\beta W_d}&=&e^{-\beta (W-\Delta F)}=\frac{Z_0}{Z_{\tau}}e^{-\beta W}= \nonumber\\
    &=& \frac{Z_0}{Z_{\tau}}\frac{e^{-\beta E_1(x_0)}}{e^{-\beta E_0(x_0)}}\frac{e^{-\beta E_2(x_1)}}{e^{-\beta E_1(x_1)}}\;...\;\frac{e^{-\beta E_{\tau}(x_{\tau-1})}}{e^{-\beta E_{\tau-1}(x_{\tau-1})}} \; .
    \label{eq:expWd_firstpassages}
\end{eqnarray}
which, toghether with Eq.\eqref{eq:E_def}, yields:
\begin{equation}
    e^{-\beta W_d} = 
    \frac{Z_0}{Z_{\tau}}\frac{Z_1\mu_1(x_0)}{Z_0\mu_0(x_0)}\frac{Z_2\mu_2(x_1)}{Z_1\mu_1(x_1)}\;...\;\frac{Z_{\tau}\mu_{\tau}(x_{\tau-1})}{Z_{\tau-1}\mu_{\tau-1}(x_{\tau-1})} = 
\frac{\mu_1(x_0)}{\mu_0(x_0)}\frac{\mu_2(x_1)}{\mu_1(x_1)}\;...\;\frac{\mu_{\tau}(x_{\tau-1})}{\mu_{\tau-1}(x_{\tau-1})} 
    \label{eq:expWd_ratiosprod}
\end{equation}
This result applies as long as the distribution $\mu_0(x)$ is stationary for the process at fixed $\lambda= \lambda_i$
and is supported on $S$, for all $i=0,\dots,\tau$. For $N$-step transformations within this framework, with $N>1$, the validity of the JE was established in \cite{hack2022jarzynski}.

Let us now take one-step evolutions going from a state $x_0$ to a state $x_1$, which means $\tau=1$ in Eq. \eqref{eq:expWd_ratiosprod}. 
Under the above assumption that the support of $\mu_0$ and $\mu_1$
is the same $\cal S$, we obtain the identity:
\begin{equation}
\left\langle e^{-\beta W_d(x)} \right\rangle = \sum_{x\in S} \frac{\mu_1(x)}{\mu_0(x)}\mu_0(x)=\sum_{x\in S}\mu_1(x)=1 \; .
    \label{eq:avexpWd_ninS}
\end{equation}
Note that the sum in Eq. \eqref{eq:avexpWd_ninS} is performed over states $x$ that belong to the state space $\mathcal{S}$ of the Markov chain associated with the \textit{initial} value $\lambda(0)=\lambda_0$ of the protocol, and that such states are eventually weighted with the distribution $\mu_1(\cdot)$, that is invariant for the \textit{final} value $\lambda(1)=\lambda_1$ of the protocol. 
In other words, the result follows from sampling the  
states $x$ belonging to the initial state space $\mathcal{S}$,
with the probability corresponding to final distribution $\mu_1(\cdot)$.
It is important to remark that the last sum in Eq. \eqref{eq:avexpWd_ninS} would not equal 1, in general, if the initial and final state spaces do not coincide.
%
%
If they do, Eq.\ \eqref{eq:avexpWd_ninS} 
shows that the JE holds 
regardless of the specific protocol. In particular, it holds even if 
the stationary states undergo a phase transition between the initial and final values of $\lambda$.
%
%
In the next subsection we will investigate variations of the ensemble averages considered above, obtained by restricting the set of observed states to a subset of the whole state space $\mathcal{S}$.

\subsection{State space truncation and Absolute Irreversibility}
\label{ss:State_Space_truncation}
\par\noindent
We now show how state space truncation, that may be due to
insufficient statistic or to finite size effects, affects the ensemble averages of the 
exponential of path-dependent observables such as $W_d$. 
We define the \textit{important state space} as the restriction  of the state space to the subset of states $\eta \subset \mathcal{S}$ that can be accessed with probability larger than a certain threshold $\bar{\mu}$
\cite{hoang2016scaling}:
\begin{equation}
\eta=\lbrace x\in \mathcal{S} \;|\; \mu(x)\geq\bar{\mu} \rbrace
    \label{eq:eta_def}
\end{equation}
One consequently has:
\begin{equation}
    \sum_{x\in\eta}\mu(x)=1-\delta
\, , \quad \mbox{where} ~~
    \delta=\sum_{x\in S \backslash \eta}\mu(x)
    \label{eq:normalization_eta}
\end{equation}
and a probability measure with support on $\eta$ is obtained  from $\mu$ dividing it by 
$1-\delta$:
\begin{equation}
\mu_{\delta}(x)=\frac{\mu(x)}{\sum_{x\in\eta}\mu(x)}=\frac{\mu(x)}{1-\delta}
    \label{eq:mu_delta}
\end{equation}
Let us now study how the average in \eqref{eq:avexpWd_ninS} is affected by this restriction of the state space to the subset $\eta$. 
Following \cite{hoang2016scaling}, we  denote $\eta_0=\lbrace x\in S \;|\; \mu_{0}(x)\geq\bar{\mu} \rbrace$ and $\eta_1=\lbrace x\in S \;|\; \mu_{1}(x)\geq\bar{\mu} \rbrace$, with same cutoff probability $\bar{\mu}$ for both $\eta_0$ and $\eta_1$.
Correspondingly, we introduce the symbols $\delta_0\equiv 1-\sum_{x\in\eta_0} \mu_0(x)$ and $\delta_1\equiv 1-\sum_{x\in\eta_1} \mu_1(x)$. Next, we call $\mathcal{I}=\eta_0 \bigcap \eta_1$, so that the sets $\eta_0= \mathcal{I} \bigcup \left(\eta_0\backslash \mathcal{I}\right)$, and $\eta_1= \mathcal{I} \bigcup \left(\eta_1\backslash \mathcal{I}\right)$ are conveniently written as the union of disjoint subsets. This way, we can write:
\begin{eqnarray}
\left\langle e^{-\beta W_d(x)} \right\rangle_{\eta_0} &=& \sum_{x\in\eta_0} \frac{\mu_1(x)}{\mu_0(x)}\mu_{0,\delta_0}(x) =
\frac{1}{1-\delta_0}\sum_{x\in\eta_0} \mu_1(x) \label{eq:final0} \; .
\end{eqnarray}
A careful rewriting of Eq. \eqref{eq:final0} yields
\begin{equation}
\left\langle e^{-\beta W_d(x)} \right\rangle_{\eta_0} 
=\frac{1-\delta_1}{1-\delta_0}+\frac{1}{1-\delta_0}\left(\sum_{x\in\eta_0\backslash\mathcal{I}} \mu_1(x)-\sum_{x\in\eta_1\backslash\mathcal{I}} \mu_1(x)\right) \;,
\label{eq:final}
\end{equation}
where we used the identity $\mathcal{I}=\eta_1\backslash(\eta_1\backslash\mathcal{I})$.
The second term in \eqref{eq:final} vanishes,  yielding 
$\langle \exp{(-\beta W_d)}\rangle_{\eta_0}=1$, even  in the presence of a reduction 
of the state space, 
if $\mu_0$ and $\mu_1$ coincide, so that $\eta_0=\eta_1$ 
and $\delta_0=\delta_1$.
The latter equality does not hold, in general, when $\mu_0\neq \mu_1$. Deviations from unity are small, as long as $\mu_0$ and $\mu_1$ differ little from one another. 
This is the case, for instance, when the initial and the final stationary states, corresponding to the initial and final values of the protocol, fall within the same macroscopic phase, as will be evidenced in the next section (see Fig. \ref{fig:avexpWdeta_vs_lambda}).

Two different cases are worth investigating when $\mu_0\neq \mu_1$.

\begin{figure}
    \centering
\includegraphics[width=0.395\textwidth]{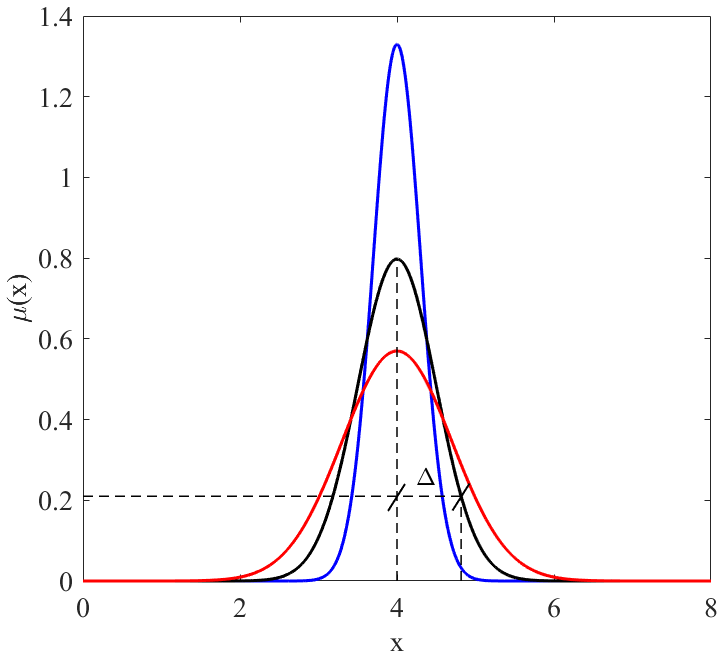} \hspace{0.25cm}
\includegraphics[width=0.395\textwidth]{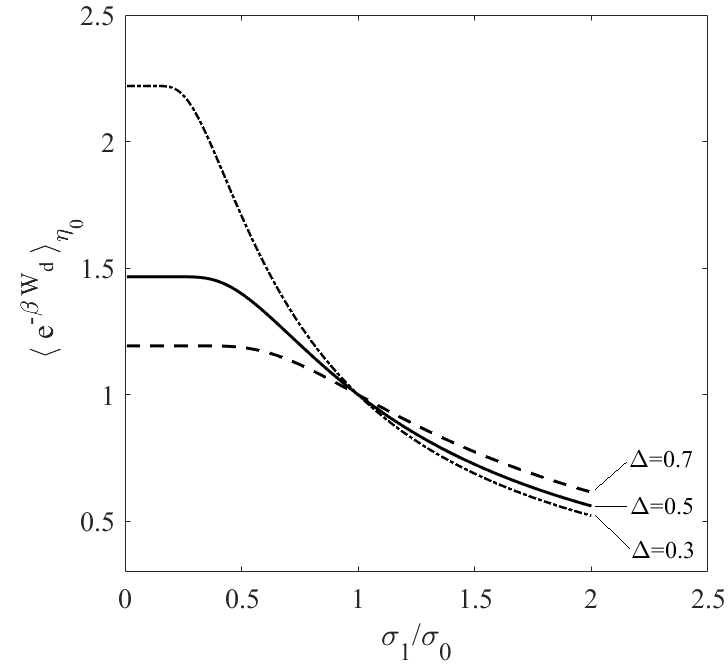} 
\includegraphics[width=0.395\textwidth]{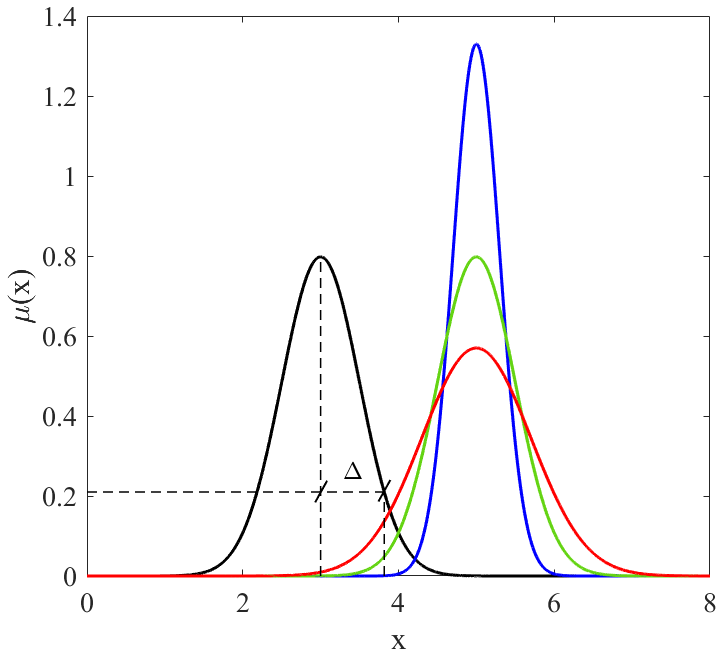} \hspace{0.4cm}
\includegraphics[width=0.415\textwidth]{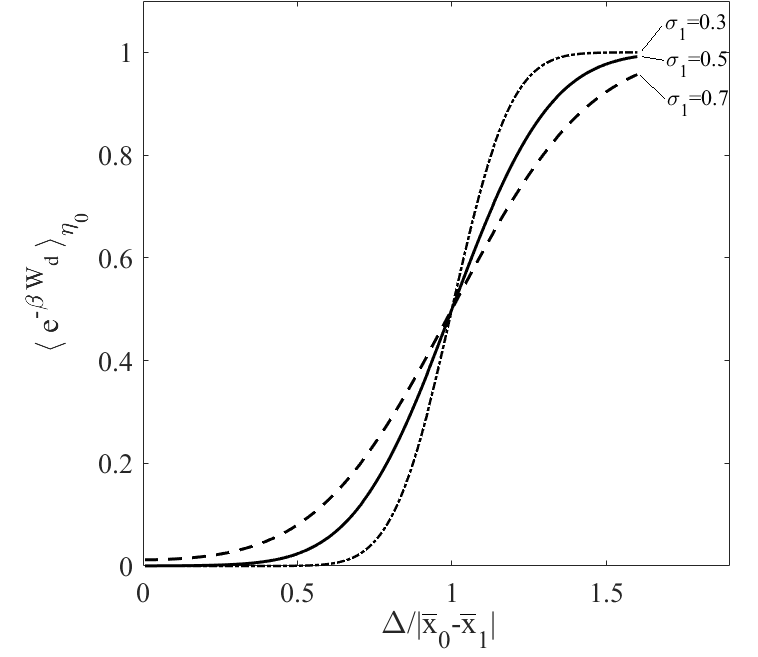}    
\caption{
\textit{Top row}: gaussian distributions centered at $\bar{x}_0=4$ (left panel):  $\mu_0(x)$ has standard deviation $\sigma_0=0.5$ (black), while the standard deviation of $\mu_1(x)$ is
$\sigma_1=0.3$ (red) and $0.7$ (blue). The horizontal dashed line denotes the value $\bar{\mu}$, which determines $\eta_0=[\bar{x}_0-\Delta,\bar{x}_0+\Delta]$. Right panel: behavior of $\langle e^{-\beta W_d}\rangle_{\eta_0}$ as a function of $\sigma_1/\sigma_0$ for different values of $\Delta$, computed using Eq. \eqref{eq:bound}.
\textit{Bottom row}, left panel:  $\mu_0(x)$ is a gaussian centered at $\bar{x}_0=3$ with standard deviation $\sigma_0=0.5$ (black curve), while $\mu_1(x)$ is a gaussian function centered at $\bar{x}_1=5$ with standard deviations $\sigma_1=$ 0.3 (blue), 0.5 (green), 0.7 (red).
Right panel: graph of $\langle e^{-\beta W_d}\rangle_{\eta_0}$ as a function of $\Delta/(\bar{x}_0-\bar{x}_1)$ for different values of $\sigma_1$, computed using Eq.\ \eqref{eq:bound}.}
\label{fig:varmu}
\end{figure}

Consider first the case with $\eta_1\subseteq\eta_0$, which implies $\mathcal{I}=\eta_1$, $\eta_1\backslash\mathcal{I}=\emptyset$ and $\eta_0\backslash\mathcal{I}=\eta_0\backslash\eta_1$. Then, 
Eq.\eqref{eq:final} yields:
\begin{eqnarray}
\left\langle e^{-\beta W_d(x)} \right\rangle_{\eta_0} 
= \frac{1-\delta_1}{1-\delta_0}+\frac{1}{1-\delta_0}\left(\sum_{x\in\eta_0\backslash\eta_1} \mu_1(x)\right) 
= \frac{1-(1-A)\delta_1}{1-\delta_0}
\label{eq:final2}
\end{eqnarray} 
where $A=\sum_{x\in\eta_0\backslash\eta_1}\mu_1(x)/\delta_1$ obeys $A\in[0,1]$. It appears that for $A$ close to $1$, the value  $\ensavg_{\eta_0}$ can exceed 1, depending on the values of $\mu_0(x)$ and $\mu_1(x)$, with $x\in\eta_0$.
For simplicity, we illustrate this fact with hypothetical gaussian distributions with standard deviations $\sigma_0$ and $\sigma_1$, centered at $x=\bar{x}$, for which analytical results are readily obtained,
cf.\ Fig.\ \ref{fig:varmu}. While such gaussians are not directly related to our finite states Markov chains, they may arise as limit cases of finite states processes, which will lead to analogous results. In the case of gaussians, the state space is  $\cal S = \mathbb R$, and the initial truncated space we consider is one interval, $\eta_0=[\bar{x}-\Delta,\bar{x}+\Delta]$, of width $|\eta_0| = 2 \Delta$.
The top right panel of the figure shows that, for different values of $\bar{\mu}$ (which is the same for both the initial and the final distributions, as generally assumed throughout this work), values of $\sigma_1/\sigma_0<1$ yield $\ensavg_{\eta_0}>1$.

Another scenario concerns distributions $\mu_0$ and $\mu_1$ for which $\eta_1\neq\eta_0$ and $\cc{I}=\emptyset$. In this case, the following bound is found:

\bea
\left\langle e^{-\beta W_d(x)} \right\rangle_{\eta_0} = \frac{1-\delta_1}{1-\delta_0}+\frac{1}{1-\delta_0}\left(\sum_{x\in\eta_0} \mu_1(x)-\sum_{x\in\eta_1} \mu_1(x)\right) 
= \frac{1}{1-\delta_0}\sum_{x\in\eta_0} \mu_1(x) < 1  \label{eq:bound}
\eea
The inequality follows from the fact that $\mu_1< \bar{\mu}$ on $\eta_0$, because $\mathcal{I}=\emptyset$, hence
$$
\sum_{x\in\eta_0} \mu_1(x)  < |\eta_0| \bar{\mu} \le \sum_{x\in\eta_0} \mu_0(x)=1-\delta_0
$$ 
where $|\eta_0|$ now represents the cardinality of $\eta_0$.
It is useful to comment on the behavior of the quantity $\langle e^{-\beta W_d}\rangle_{\eta_0}$ as a function of $|\eta_0|$, whose value can be tuned by varying the cutoff probability $\bar{\mu}$. For simplicity, we refer again to gaussian distributions, taking 
$\eta_0 =[\bar{x}-\Delta,\bar{x}+\Delta]$, and
$\bar{\mu}$ and $\Delta$ in a range that allows $\mathcal{I}=\emptyset$,
so that Eq.\ \eqref{eq:bound} applies. In the bottom left panel of Fig.\ \ref{fig:varmu} the distributions $\mu_0$ and $\mu_1$ are gaussian densities centered at $\bar{x}_1$ and $\bar{x}_1 \neq \bar{x}_0$, with standard deviations $\sigma_0$ and $\sigma_1$, respectively.
The bottom right panel of Fig.\ \ref{fig:varmu} shows that, for  values of $\Delta$ guaranteeing $\mathcal{I}=\emptyset$, a decrease of $\Delta$ leads to a decrease of $\langle e^{-\beta W_d}\rangle_{\eta_0}$ to values substantially smaller than 1.

One possible mechanism leading to a reduction of the available state space, is known as \textit{absolute irreversibility} \cite{murashita2014nonequilibrium}, which is also investigated within the realm of Ising models 
\cite{hoang2016scaling}. 
Essentially, absolute irreversibility occurs if the probability of a certain ``forward'' path in the state space is zero, whereas the probability of the corresponding ``reverse'' path (tracing back in opposite order all the states visited along the forward bath) is non-zero. This phenomenon is experimentally realized when an external field keeps the system first trapped inside a subregion of the state space, and then the constraint is released and the system can explore the whole state space.
This is also the case of {\em e.g.}\ the free expansion of a gas, in which particles are initially confined by a wall inside a fraction of the volume of the box and then, after the swift removal of the wall, they can eventually fill up the entire volume. We denote by $\mu_1(x)$ and $\mu_0(x)$ the probability distributions describing the final and initial states, respectively, that are supported on $\mathcal{S}_1$ and $\mathcal{S}_0 \subset \mathcal{S}_1$. In such cases, Eq.\ \eqref{eq:avexpWd_ninS} yields:
\begin{eqnarray}
\left\langle e^{-\beta W_d(x)} \right\rangle 
&=&\sum_{x\in\mathcal{S}_0} \mu_1(x) < 1 \label{eq:absirr} \; .
\end{eqnarray}

The key aspect of these processes -- which encodes much of the physics of the model -- is that the state space in the initial state is intrinsically smaller than that in the final state. At times, absolute irreversibility and the state space truncation appear to be interwoven. In Sec. \ref{s:mod1} we shall discuss a model in which the two phenomena in fact mingle, while $\mathcal{I}=\emptyset$. In this case, one finds that
\begin{equation}
  \left\langle e^{-\beta W_d(x)} \right\rangle_{\eta_0} <  \left\langle e^{-\beta W_d(x)} \right\rangle<1 \label{eq:combin} \; ,
\end{equation}
where the first inequality holds provided that  
\begin{equation*}
\sum_{x \in \mathcal{S}_0\backslash\eta_0}\mu_0(x)\sum_{x \in \eta_0}\mu_1(x)< \sum_{x \in \eta_0}\mu_0(x) \sum_{x \in \mathcal{S}_0\backslash\eta_0}\mu_1(x) 
\end{equation*} Note, indeed, that the latter inequality can be rewritten as
$$
0 < -\delta_0 \sum_{x \in \eta_0}\mu_1(x)+(1-\delta_0) \sum_{x \in \mathcal{S}_0\backslash\eta_0}\mu_1(x)
$$
Next, adding $\sum_{x \in \eta_0}\mu_1(x)$ on both sides, one obtains
$$
\sum_{x \in \eta_0}\mu_1(x) < (1-\delta_0)\left( \sum_{x \in \eta_0}\mu_1(x)+\sum_{x \in \mathcal{S}_0\backslash\eta_0}\mu_1(x)  \right)
$$
hence 
$$
\frac{1}{1-\delta_0} \sum_{x \in \eta_0}\mu_1(x) < \sum_{x \in \mathcal{S}_0}\mu_1(x) 
$$
which yields the result. 

\section{The models}
\label{s:models}
\par\noindent

In this section we investigate different models for which an explicit analytical solution can be provided. The Jarzynski relation, revisited with the notions of state space reduction and absolute irreversibility introduced in Sec.\ \ref{ss:State_Space_truncation}, are used to detect possible  discontinuous behaviors of an order parameter as a function of an external control parameter. 
First, a probabilistic urn model, described by a Markov chain, is considered, in which a threshold parameter allows to tune between different steady states. The second model is the classical 2D  Ising model subject to an external magnetic field. 

\subsection{The urn model}
\label{s:mod1}

\begin{figure}
\subfigure{\includegraphics[width=0.25\textwidth]{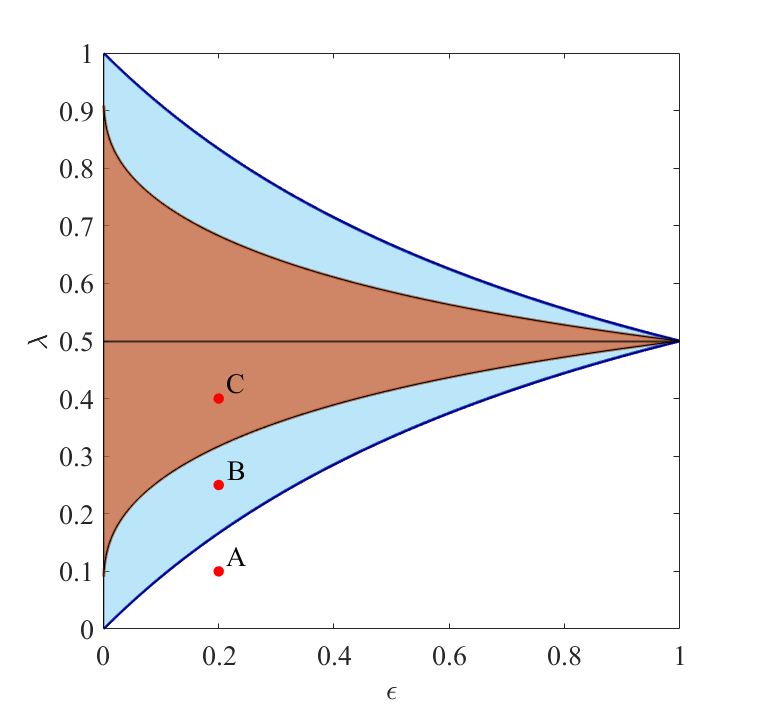}} 
\subfigure{\includegraphics[width=0.24\textwidth]{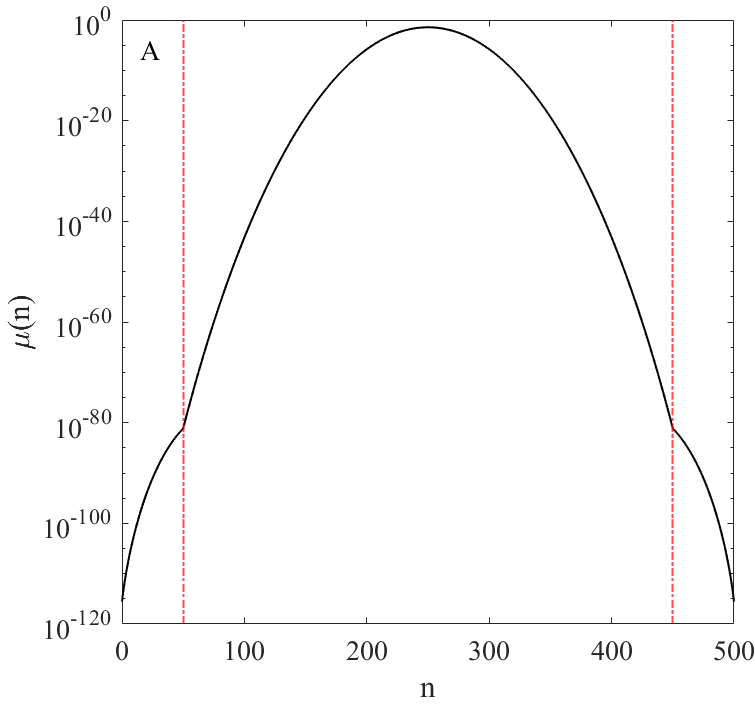}} 
\subfigure{\includegraphics[width=0.24\textwidth]{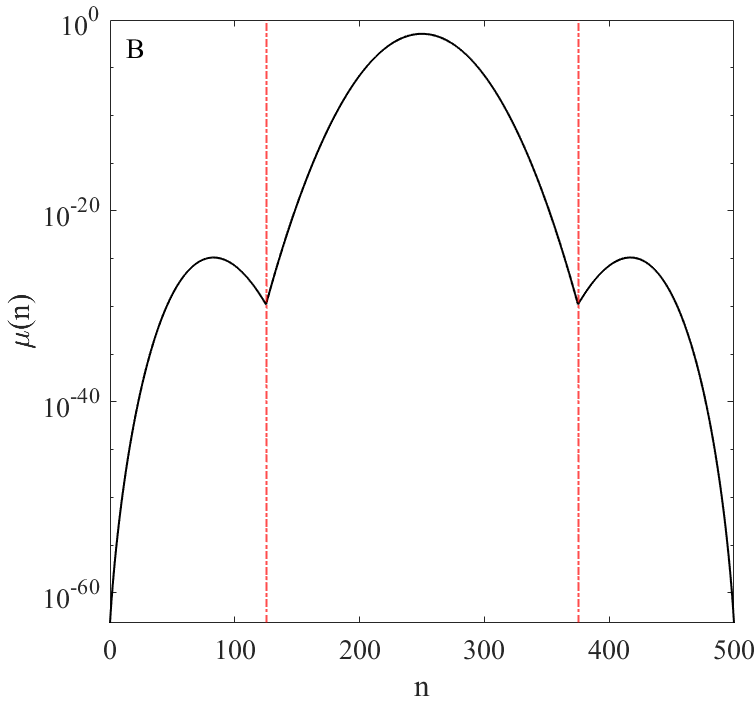}} 
\subfigure{\includegraphics[width=0.24\textwidth]{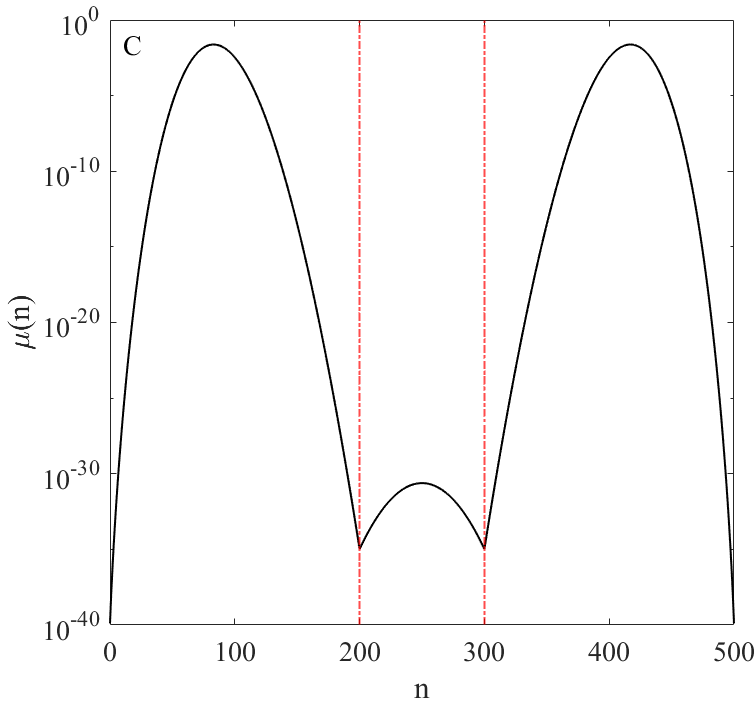}}
\caption{Stationary distributions for $N=500$.
Right panel: stability regions in the $\epsilon$-$\lambda$ parameter space. 
White corresponds to stable equilibrium ($n=N/2$); cyan to two metastable nonequilibrium states $(n=n^* , N-n^*)$ and stable equilibrium; brown to 
two stable nonequilibrium states and metastable equilibrium.
Remaining three panels, from left to right, plot with vertical $\log$-scale
the stationary distributions
\eqref{eq:mut_TlessNhalf}
referring to the points 
$A=(0.2 , 0.1)$, $B=( 0.2 , 0.25 )$ and 
$C= (0.2 , 0.4)$.}   
\label{fig:parameter_space_epsilonlambda}
\end{figure}

The modification of the classical Ehrenfest urn model \cite{Kac,Feller} that we analyze 
was originally introduced in \cite{cirillo2020deterministic} to understand billiard dynamics comprising a Maxwell demon  \cite{cirillo2021deterministic2,cirillo2022transport,cirillo2023deterministic}. 
It refers to $N$ particles initially distributed inside two urns. At each discrete time step one of the two urns is selected with a probability proportional to the number of particles contained in it, and then one of such particles is moved to the other urn, unless a clogging phenomenon stops it. To represent this process,
let $n(t)\in \mathcal{S}=\lbrace0, 1, . . . , N\rbrace$ denote the number of particles in one selected urn at time $t\in \mathbb{N}$, 
and introduce a parameter
$\lambda\in[0,1]$, to define a
threshold $T=\lfloor\lambda N\rfloor \in \lbrace 0,1,...,N\rbrace$ which implements
the clogging mechanism together with another parameter
$\epsilon \in [0,1]$. We then consider the 
discrete time Markov chain $\lbrace n(t): t\ge 0 \rbrace$
that obeys the following transition rules:
\begin{equation}
	p^{(t)}_{n,n+1}(T)=\frac{N-n}{N}\epsilon_1(T), \;\;\; p^{(t)}_{n,n-1}(T)=\frac{n}{N}\epsilon_2(T)
	\label{eq:trans_prob}
\end{equation}
where
\begin{align}
	\epsilon_1(T)=
	\begin{cases}
	\epsilon, &  n<N-T \\
	1, &  n\geq N-T
	\end{cases} 
	,\;\;\;\;\;\epsilon_2=
	\begin{cases}
	1, & n\leq T \\
	\epsilon, & n>T
	\end{cases}
	\label{eq:eps1_eps2}
\end{align}
The stationary distributions $\mu_{\lambda}(n)$ of the process take the form \cite{cirillo2020deterministic}:
    \begin{equation}
    \mu_{\lambda}(n)=\frac{1}{C_{\lambda}}\binom{N}{n}
        \begin{cases}
            \epsilon^n & \textit{if} \hspace{0.5cm} n\leq T \\
            \epsilon^T & \textit{if} \hspace{0.5cm} T+1\leq n\leq N-T-1 \\
            \epsilon^{N-n} & \textit{if} \hspace{0.5cm} n\geq N-T
        \end{cases}
        \label{eq:mut_TlessNhalf}
    \end{equation}
    when $T < N/2$, and
    \begin{equation}
    \mu_{\lambda}(n)=\frac{1}{C_{\lambda}}\binom{N}{n}
        \begin{cases}
            \epsilon^n & \textit{if} \hspace{0.5cm} n\geq N-T+1 \\
            \epsilon^{N-T+1} & \textit{if} \hspace{0.5cm} N-T+2\leq n\leq T-1 \\
            \epsilon^{N-n} & \textit{if} \hspace{0.5cm} n\leq T 
        \end{cases}
        \label{eq:mut_TgreaterNhalf}
    \end{equation}
when $T>N/2$, where $C_{\lambda}$ is the normalization coefficient associated to $\mu_{\lambda}$. We only analyze the case $T<N/2$,
as the case $T>N/2$ trvially follows. The dependence of the stationary state on the parameters $\epsilon$ and $\lambda$, along with the stationary distribution $\mu(n)$ evaluated in different regions of the parameter space are shown in 
Fig. \ref{fig:parameter_space_epsilonlambda}, showing that $\mu$ concentrates about different states for different values $\lambda$. In particular, the left panel of 
Fig. \ref{fig:parameter_space_epsilonlambda} reveals the presence, in  parameter space, of equilibrium as well as nonequilibrium phases, portrayed as white and brown regions, respectively \cite{cirillo2020deterministic}. 
While in the equilibrium region the stationary distribution peaks at $n=N/2$ (second panel from left), in the nonequilibrium ones the distribution concentrates
around the two states $n=n^*=(\varepsilon N-1)/(\varepsilon+1)$ and $n=N-n^*$ (right panel). Moreover, in the light blue region, shown in the left panel, the states $n=n^*$ and $n=N-n^*$ are metastable while $n=N/2$ is stable. 
The distribution corresponding to the latter case shows a global maximum at $n=N/2$ and two symmetric local maxima at $n=n^*$ and $n=N-n^*$ (third panel). 

\begin{figure}
\centering  \subfigure{\includegraphics[width=0.35\textwidth]{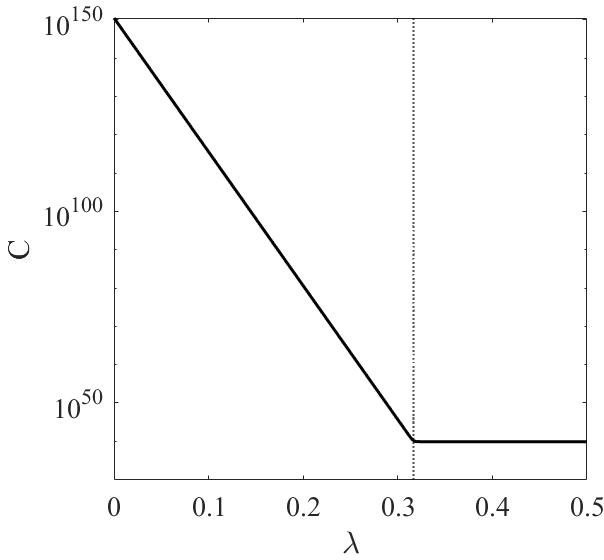}} \hspace{1cm} \subfigure{\includegraphics[width=0.33\textwidth]{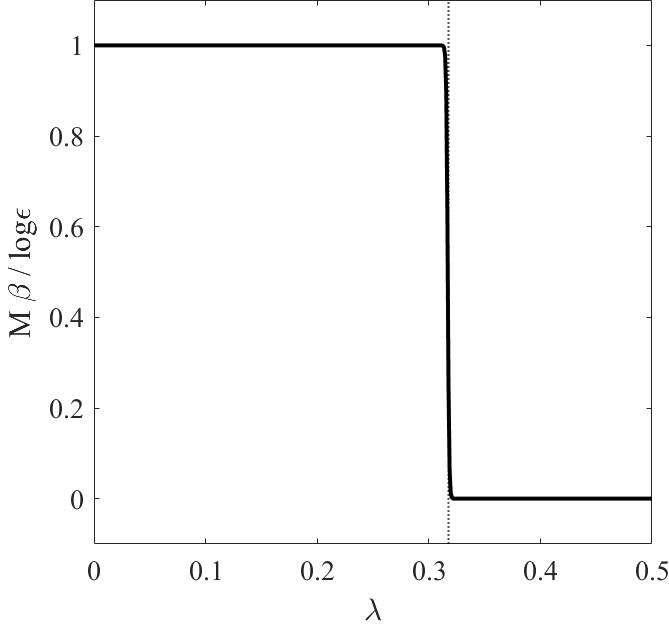}}
\caption{Left panel: Normalization constant $C_\zl$ as a function of $\lambda$, Eq. \eqref{eq:C_t}, with vertical axis in logarithmic scale. Dotted vertical line indicates the critical value $\lambda^*$ defined in Eq. \eqref{eq:crit_lambda}. Right panel: $M(\lambda)\beta/\log{\epsilon}$ obtained through Eq. \eqref{eq:M_explicit}, with  dotted vertical line  indicating the critical value $\lambda^*$. Both plots are obtained for $N=500$, $\beta=1$ and $\epsilon=0.2$.}
    \label{fig:C_vs_T}
\end{figure}

Let us now calculate the normalization coefficient $C_{\lambda}$ and analyze its dependence on the threshold parameter to highlight its singular behavior. First, observe that
\bea
C_{\lambda} &=& \sum_{n=0}^T \epsilon^n \binom{N}{n} + \sum_{n=T+1}^{N-T-1} \epsilon^T \binom{N}{n} + \sum_{n=N-T}^N \epsilon^{N-n} \binom{N}{n} 
= 2\sum_{n=0}^T \epsilon^n \binom{N}{n} + \sum_{n=T+1}^{N-T-1} \epsilon^T \binom{N}{n} \nonumber \\[6pt]
& \approx & 2\sum_{n=0}^N \epsilon^n \binom{N}{n} + \sum_{n=0}^{N} \epsilon^T \binom{N}{n} 
= 2(1+\epsilon)^N + 2^N \epsilon^T 
\label{eq:C_t}
\eea
where the last equality follows from the binomial theorem. Therefore, for $\epsilon<1$ and $T > 1$, 
$C_{\lambda}$ is an exponentially decaying function of $\lambda$. Moreover, the term $(1+\epsilon)^N$ is negligible with respect to $2^N\epsilon^T$ if
\be
T < N\log\left(\frac{1+\epsilon}{2}\right)(\log\epsilon)^{-1} \nonumber
\ee
while 
$2^N\epsilon^T$ is negligible
with respect to 
$(1+\epsilon)^N$ if 
\be
T > N\log\left(\frac{1+\epsilon}{2}\right)(\log\epsilon)^{-1} \nonumber
\ee
In the large $N$ limit, the separation between these two regimes becomes sharp, 
and defines the critical value \cite{cirillo2020deterministic}
\be
\lambda^*=\log\left(\frac{1+\epsilon}{2}\right)(\log\epsilon)^{-1} 
\label{eq:crit_lambda}
\ee
of a phase transition, implying:
\begin{eqnarray}
C_{\lambda<\lambda^*}&\approx&2^N\epsilon^T, \nonumber\\
C_{\lambda>\lambda^*}&\approx&2(1+\epsilon)^N.
\label{eq:CT_approx}
\end{eqnarray}
The behavior of $C_\zl$ as a function of the normalized threshold $\lambda$ is illustrated in Fig. \ref{fig:C_vs_T}.

Mimicking the theory of phase transitions, let us now look at the first derivative with respect to $\zl$ of the quantity $\mathcal{F} = -\beta^{-1}\log{C}$:
\be
M \coloneqq -\frac{1}{N}\frac{\partial}{\partial\lambda}\mathcal{F}=-\frac{1}{N}\frac{\partial}{\partial\lambda}\left(-\frac{1}{\beta}\log{C(\lambda)}\right)
\label{eq:M}
\ee
which using Eq.\eqref{eq:C_t},
and taking large $N$ so that  $T\approx\lambda N$, writes:
\be
M=\frac{\log{\epsilon}}{\beta}\frac{2^N\epsilon^{\lambda N}}{(1+\epsilon)^N+2^N\epsilon^{\lambda N}}=\frac{\log{\epsilon}}{2\beta}\left[1+\tanh{\left(\frac{N}{2}\left( \lambda-\lambda^*\right)\log\epsilon\right)}\right]
\label{eq:M_explicit}
\ee
The graph of $M(\lambda)$, which acts as an order parameter for this model, is shown in the right panel of Fig. \ref{fig:C_vs_T} for $N=500$. At such moderate $N$, hence even more sharply in the large $N$ limit, $M$ undergoes a jump discontinuity at $\lambda=\lambda^*$, which in statistical mechanics is typically associated with first order phase transitions.

\begin{figure}
    \centering   \subfigure{\includegraphics[width=0.35\textwidth]{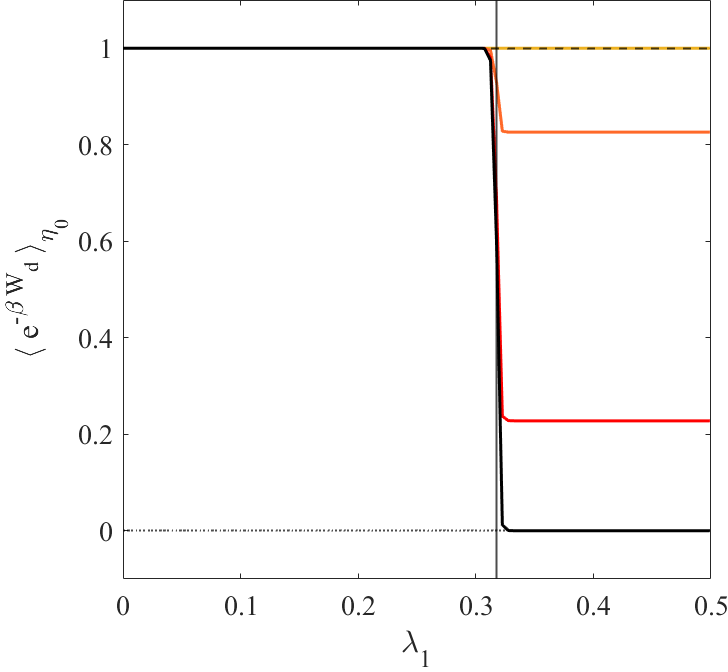}} \hspace{1cm} \subfigure{\includegraphics[width=0.35\textwidth]{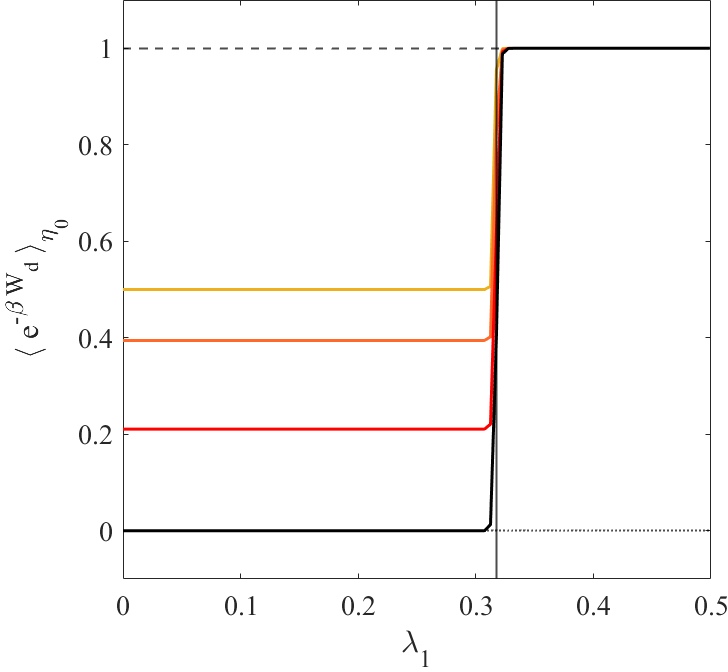}}
\caption{
Numerical estimation of $\langle e^{-\beta W_d}\rangle_{\eta_0}$, as a function of $\lambda_1$,
given by Eq.\ \eqref{eq:final}.
The yellow, orange, red and black lines correspond to cutoffs 
$\bar{\mu}=0,\;10^{-60},\;10^{-50},\;10^{-20}$ respectively. The vertical solid black lines lie at $\lambda=\lambda^*$. The dashed black lines correspond to Eq.\ \eqref{eq:avexpWd_ninS}. $N$, $\beta$ and $\epsilon$ are set to the same values considered in Fig. \ref{fig:C_vs_T}.
\textit{Left panel}: The process starts in a homogeneous state with $\lambda_0=0$. The yellow line is always at 1; the phase transition is only revealed by the truncated distributions.
\textit{Right panel}: the process starts in a non-homogeneous state 
with $\lambda_0 = 0.5$.
The values for $\bar{\mu}=0$ and $\lambda<\lambda^*$ do not equal 1, as prescribed by Eq.\ \eqref{eq:avexpWd_ninS2}. Even the full distribution reveals the phase transition.
}
\label{fig:avexpWdeta_vs_lambda}
\end{figure}

Suppose now that the threshold depends on time, so that
$\lambda=\lambda(t)$, with $t = 0,1$, constitutes a protocol in accord with the theory illustrated in Sect.\ \ref{WFEMC}, and $\mu_{\lambda(t)}$ is understood as the stationary distribution corresponding to a \textit{fixed} value of $t$ \cite{hack2022jarzynski}.
In particular, take the one-step transformation expressed by the protocol $\lambda_0\rightarrow \lambda_1$, with $\lambda_0\equiv\lambda(0)=0$ and $\lambda_1\equiv\lambda(1) \in [0 ,  0.5]$. The associated stationary distributions 
can be computed using Eq. \eqref{eq:mut_TlessNhalf} and \eqref{eq:CT_approx}. Specifically, denoting by $\mu_0(n)\equiv\mu_{\lambda_0}(n)$ the distribution evaluated at $\lambda=\lambda_0$, we find
\be
\mu_0(n)=\frac{1}{2^N}\binom{N}{n} \; , \nonumber
\ee
whereas $\mu_1(n)\equiv\mu_{\lambda_1}(n)$ depends on the final value $\lambda_1$ of the protocol.
As discussed earlier in Sec. \ref{ss:State_Space_truncation}, a summation over all possible initial states $n\in \mathcal{S}_0 = \lbrace0,1,...,N\rbrace$ in Eq. \eqref{eq:final}, which corresponds to no state space reduction ($\bar{\mu}=0$), yields the JE, regardless of the chosen value of $\lambda_1$. Specifically, for $\lambda^*<\lambda_1\le 0.5$ two symmetric steady states $n=n^*$ and $n=N-n^*$ appear. Indeed, we have $\mathcal{S}_0= \mathcal{S}_1 = \{ 0,1,\dots,N \}$, and Eq.\ \eqref{eq:avexpWd_ninS} applies,
independently of $\zl_1$.
Therefore, the JE does not reveal the jump discontinuity portrayed in the right panel of Fig. \ref{fig:C_vs_T}, because it is not affected by that. 
The situation is different if the state space is reduced to a subset $\eta_0$ by a cutoff probability $\bar{\mu} > 0$, see Eq. \eqref{eq:final0}. 
The left panel of Fig. \ref{fig:avexpWdeta_vs_lambda} shows indeed that $\ensavg_{\eta_0}\simeq 1$ for $\lambda_1<\lambda^*$, while a sudden drop to a value $\ensavg_{\eta_0} < 1$ occurs for $\lambda_1>\lambda^*$. 
In fact, $\bar{\mu}$ as small as $10^{-20}$ suffices to get $\ensavg_{\eta_0}\simeq 0$ for $\lambda_1>\lambda^*$. This behavior signals the sudden separation of the state space regions occupied by the sets $\eta_0$ and $\eta_1$, in Eq. \eqref{eq:final}, occurring at $\lambda_1=\lambda^*$. While for $\lambda_1<\lambda^*$ the sets $\eta_0$ and $\eta_1$ overlap consistently, for $\lambda_1>\lambda^*$ one can safely consider $\mathcal{I}=\emptyset$, provided $\bar{\mu}$ is large enough. 
Thus, following the discussion of Sec. \ref{ss:State_Space_truncation}, for $\lambda_1>\lambda^*$ 
one expects not only that $\ensavg_{\eta_0}<1$, but also that, this average decreases
for growing $\bar{\mu}$. This is indeed demonstrated in the left panel of Fig.\ \ref{fig:avexpWdeta_vs_lambda}.

Consider now the one-step transformations with  $\lambda_0=0.5$, which is larger than $\zl^*$, and any $\lambda_1 \in [0,1]$.
In this case, the translational symmetry is broken, because 
the initial condition selects only one of the two nonequilibrium states, say the one with $n=n^*$.
Then, taking $T=N/2$ in Eq.\ \eqref{eq:mut_TlessNhalf} yields \begin{equation*}
\mu_{\zl=0.5}=\frac{1}{C_{0.5}}\binom{N}{n}\epsilon^n, \;\; n<N/2 \;.
\end{equation*}
where the second row of Eq.\ \eqref{eq:CT_approx} turns $C_{0.5} \approx (1+\epsilon)^N$,
because only half of the state space $\mathcal{S}$ is involved. 
One obtains:
\begin{eqnarray*}
\mu_{\zl(0)}(n)=
\begin{cases}
\frac{\epsilon^n}{(1+\epsilon)^N}\binom{N}{n} \;\; \text{for} \;\; n< N/2 \\
0 \;\; \text{for} \;\; n\geq N/2
\end{cases}
\end{eqnarray*}
which is supported on $\mathcal{S}_0=\{0,\dots,N/2-1\}$, while the support of the distribution $\mu_{\zl(1)}$ is $\mathcal{S}_1=\mathcal{S}_0$ for $\lambda^*<\lambda_1\le0.5$ (nonequilibrium phase) and $\mathcal{S}_1=\{0,\dots,N\}$ for $0\le \lambda_1 <\lambda^*$ (equilibrium phase). Consequently, Eq.\ \eqref{eq:avexpWd_ninS} holds for $\lambda^*<\lambda_1\le0.5$ whereas Eq.\ \eqref{eq:absirr} applies for $0\le \lambda_1 <\lambda^*$, because $\mathcal{S}_0\subset \mathcal{S}_1$, which implies:

\begin{eqnarray}
\left\langle e^{-\beta W_d(n)} \right\rangle &=&\sum_{n\in\mathcal{S}_0}\mu_1(n)
= \sum_{n=0}^{T-1}\frac{\epsilon^n}{2^N\epsilon^T}\binom{N}{n} + \sum_{n=T}^{N/2-1}\frac{1}{2^N}\binom{N}{n} \nonumber \\
&\simeq& \sum_{n=T}^{N/2-1}\frac{1}{2^N}\binom{N}{n}  \simeq  1/2 \; , 
    \label{eq:avexpWd_ninS2} 
\end{eqnarray}
where the approximations turn exact for $T=0$ and $N\rightarrow\infty$ . The violation of the JE stems from the ``absolute irreversibility''
of the process $\zl_0 \to \zl_1$. This is illustrated by the numerical results reported in the right panel of Fig. \ref{fig:avexpWdeta_vs_lambda}, which also clarifies the effect of reducing the state space with a cutoff probability $\bar{\mu}$. The effect is a $\bar \mu$ dependent further decrease of the average of $e^{-\beta W_d}$ for $0\le \lambda_1 <\lambda^*$, visible in the left panel of Fig.\ \ref{fig:avexpWdeta_vs_lambda}, cf.\ Eq.\ \eqref{eq:combin}.

\begin{figure}
    \centering    \subfigure{\includegraphics[width=0.35\textwidth]{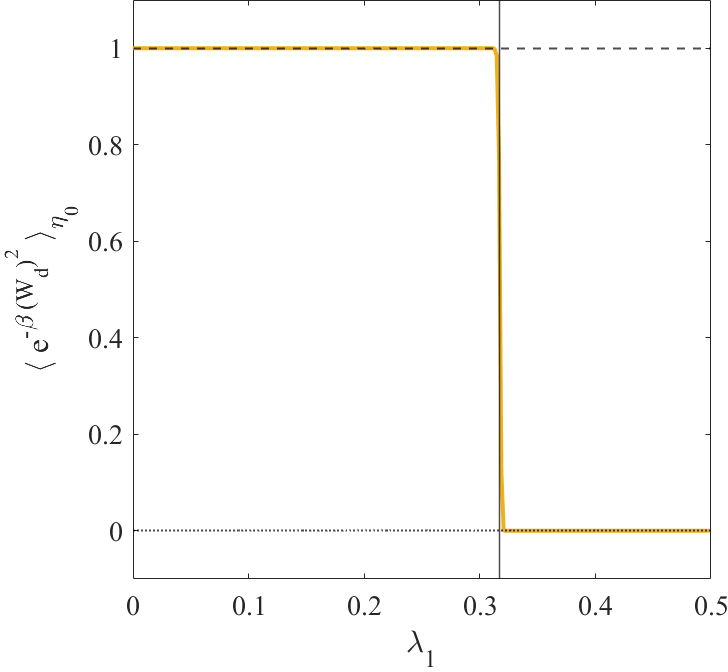}}
    \hspace{1cm} \subfigure{\includegraphics[width=0.35\textwidth]{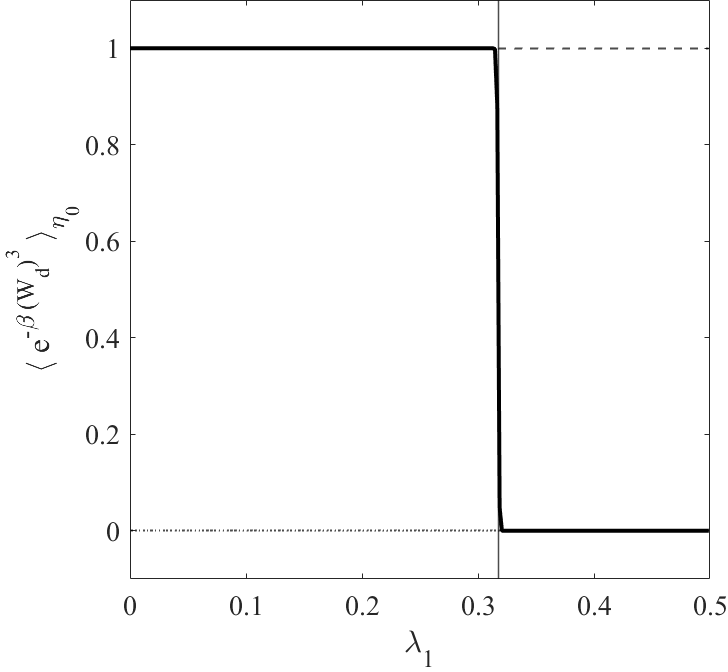}} \\
    \subfigure{\includegraphics[width=0.35\textwidth]{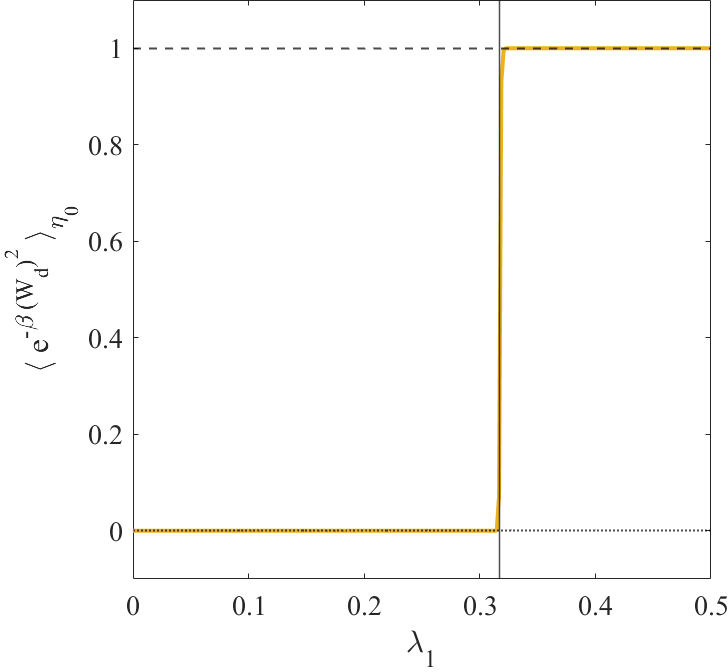}}
    \hspace{1cm} 
 \subfigure{\includegraphics[width=0.35\textwidth]{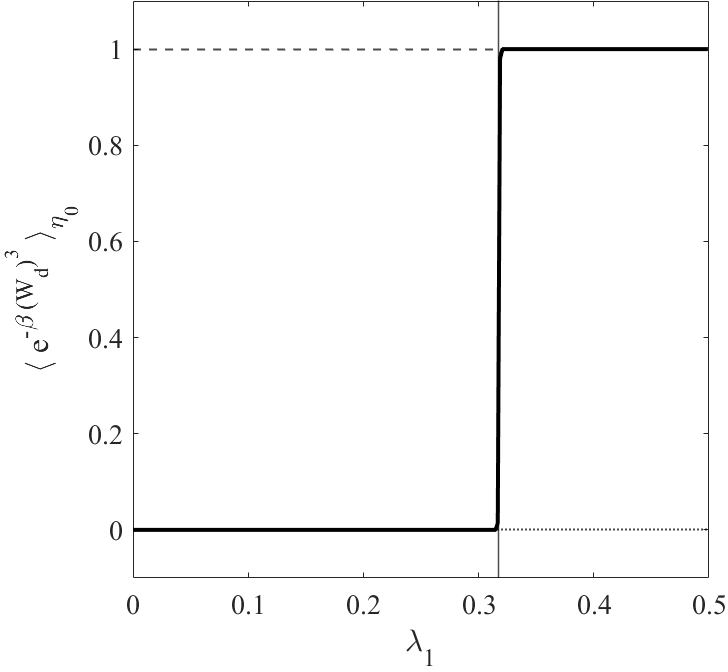}}
    
\caption{Numerical estimation of $\langle e^{-\beta (W_d)^p}\rangle_{\eta_0}$ as a function of $\lambda_1$ for  $p=2$ and $p=3$ (left and right column, respectively), with the parameters $N$, $\beta$ and $\epsilon$ set to the same values as in Fig. \ref{fig:C_vs_T}. The plots refer to a process starting in a homogeneous state with $\lambda_0=0$ (top row) or in a non-homogeneous state with $\lambda_0 = 0.5$ (bottom row).
In the left panels, the yellow line refers to the cutoff  $\bar{\mu}=0$, whereas the curves corresponding to cutoff values $\bar{\mu}>0$ are not shown in the plots, as they are graphically indistinguishable from the yellow line. In the right panels, the solid black line refers to the cutoff $\bar{\mu}=10^{-15}$.}

\label{fig:avexpWdeta_tothep_vs_lambda}
\end{figure}

We conclude this Section observing that different exponential 
variables may be used to efficiently identify phase transitions. For instance, we considered $\exp\{-\beta \left( W_d\right)^p\}$, with $p\in\mathbb{N}$, and numerically studied the cases $p=2$ and $p=3$. The numerical evaluation of $\langle \exp\{-\beta \left( W_d\right)^p\} \rangle_{\eta_0}$ for the two different values of $p$ and for different initial values of $\lambda_0$ is represented in Fig. \ref{fig:avexpWdeta_tothep_vs_lambda}.

It is important to note that, unlike the standard expression,
corresponding to $p=1$, the quantity $\langle e^{-\beta \left( W_d\right)^2} \rangle_{\eta_0}$ displays a sharp transition at $\lambda=\lambda^*$ even for $\bar{\mu}=0$, and not only for the truncated distributions. Moreover, $\left( W_d\right)^2$ is a quantity substantially different from $W_d$, because it only takes non-negative values, while negative values are fundamental for the usual fluctuation relations. This indicates that our approach may result effective with variables of quite different nature.

Let us also note that
taking higher odd powers of
$W_d$, the calculation of the full state space averages may become prohibitive, from a computational point of view. Indeed, the negative values of
$W_d$, raised to a large odd power and exponentiated, may 
result non-computable. However, our method based on truncated distributions may remain effective, cf.\ 
Fig. \ref{fig:avexpWdeta_tothep_vs_lambda} right panel, for $p=3$.

\FloatBarrier

\subsection{The 2D Ising Model}
\label{ss:ising}

The second example we tackle is the classical 2D Ising model under an external magnetic field $h$, defined on a $2$D lattice $\Lambda=\{1,\dots,L\}^2$, with  the state space 
$\mathcal{S}=\{-1,+1\}^{\Lambda}$. Here $\beta=(k_B T)^{-1}$ denotes the inverse temperature.
Let $\mathbf{\sigma}=(\sigma_1,\dots,\sigma_{|\Lambda|})\in \mathcal{S}$ denote a generic \textit{spin configuration}, and introduce the Ising Hamiltonian, which reads
$$ 
H(\sigma)=-\frac{1}{2} J \sum_{\substack{\langle j,k\rangle \\ j,k \in \Lambda}} \sigma_j \sigma_k
-h \sum_{j\in \Lambda}\sigma_j \; ,
$$
where $\sigma_j=\{-1,+1\}$ for $j\in \Lambda$, $J>0$ denotes the ferromagnetic coupling constant and the first sum is taken over all nearest neighbor spins. Periodic boundary conditions are imposed along the horizontal and vertical directions. 
 The model exhibits a critical value  $\beta_c=\ln{(1+\sqrt{2})}/2J$, marking the transition from a disordered (paramagnetic) to an ordered (ferromagnetic) phase. We then let the the 
magnetic field depend on time, defining a protocol $\zl(t)=h(t)$.
In Ref.\ \cite{hoang2016scaling},
a similar problem is investigated, taking $\lambda(t)=J(t)$. 
We denote by $H_i(\sigma)$ and $H_f(\sigma)$ the Hamiltonians evaluated at the initial and the final values of the protocol, respectively, and we take $\mu_k(\sigma)=\exp{(-\beta H_k(\sigma))}/Z_i$ as the Gibbs measure at the inverse temperature $\beta$, with 
$k=0,1$ and
$Z_k=\sum_{\sigma \in \mathcal{S}}\exp{(-\beta H_k(\sigma))}$. So both $\mu_0$ and $\mu_1$ are supported on 
$\mathcal{S}$.

We now introduce the reduced state spaces $\eta_0=\{\sigma|\mu_0(\sigma)\ge \bar{\mu}\}$ and $\eta_1=\{\sigma|\mu_1(\sigma)\ge \bar{\mu}\}$ and we correspondingly denote $\overline{Z}_0=\sum_{\sigma \in \eta_{0}}\exp{(-\beta H_0(\sigma))}$ and $\overline{Z}_1=\sum_{\sigma \in \eta_{1}}\exp{(-\beta H_1(\sigma))}$. With the notation of Sec.\ \ref{ss:State_Space_truncation}, we have:

\begin{equation}
1-\delta_k =\sum_{\sigma\in\eta_{k}}\mu_k(\sigma)=\frac{\sum_{\sigma\in\eta_k} \exp{(-\beta H_k(\sigma))}}{\sum_{\sigma\in\mathcal{S}} \exp{(-\beta H_k(\sigma))}} =\frac{\overline{Z}_k}{Z_k} \;,
\quad k=0,1 \;,
\end{equation}
which yields:
\begin{eqnarray*}
  \frac{\overline{Z}_0}{\overline{Z}_1}=\frac{Z_0}{Z_1}\frac{1-\delta_0}{1-\delta_1} \; .  
\end{eqnarray*}
Furthermore, for a function $f(\sigma):\mathcal{S}\rightarrow \mathbb{R}$, we set: 
\begin{equation}
\langle f \rangle_{\eta_0}=\sum_{\sigma\in\eta_{0}} f(\sigma) \frac{e^{-\beta H_0(\sigma)}}{\overline{Z}_0}=\frac{1}{1-\delta_0}\sum_{\sigma\in\eta_0} f(\sigma) \mu_0(\sigma) \; .
\label{avegi}
\end{equation}
Then, taking
$$
f(\sigma)= e^{-\beta W_d}=e^{-\beta (W-\Delta F)}=e^{-\beta (H_1(\sigma)-H_0(\sigma))}\frac{Z_0}{Z_1} \; ,
\quad \mbox{where} ~~ 
\beta \Delta F=\log \cfrac{Z_0}{Z_1} \; ,
$$ 
we find 
$$
f(\sigma) \mu_0(\sigma)=\frac{e^{-\beta H_1(\sigma)}}{Z_0}\frac{Z_0}{Z_1}=\mu_1(\sigma) \;,
$$
and
\begin{eqnarray}
    \left\langle e^{-\beta W_d}\right\rangle_{\eta_0}=\frac{1}{1-\delta_0}\sum_{\sigma\in\eta_{0}} \mu_1(\sigma) \; .
    \label{eq:finalIsing}
\end{eqnarray}
The latter equation coincides with Eq.\ \eqref{eq:final} and recovers \cite[Eq.\ $14$]{hoang2016scaling}.

We are interested in studying the application of Eq.\
\eqref{eq:finalIsing} to evolutions in which the magnetic field changes from $\lambda_0 = h_0$ to $\lambda_1=h_1$.

Setting $J=1$, which yields $\beta_c\approx 0.4406$, we consider different scenarios. We first choose  $\beta>\beta_c$, so that the 
2D Ising model undergoes a first order phase transition when the magnetic field $h$ changes sign, and we study the behavior of $\langle e^{-\beta W_d} \rangle_{\eta_0}$ as a function of $h_1$. Later, we study the case with $\beta<\beta_c$, where no phase transition takes place. 
The computation of $\langle e^{-\beta W_d} \rangle_{\eta_0}$ can be conveniently carried out by 
sampling configurations $\sigma\in \mathcal{S}$, with a prescribed magnetization density $m=N^{-1}\sum_{j\in\Lambda}\sigma_j$, from the distribution \cite{hoang2016scaling}:
\begin{equation*}
P(m)=\sum_{\sigma\in \mathcal{S}}\mu(\sigma) \, \delta\left(m-\frac{1}{N}\sum_{j\in\Lambda}\sigma_{j\in \Lambda}\right) 
\end{equation*}

\begin{figure}
        \centering \subfigure{\includegraphics[trim = {3cm 9.5cm 3cm 9.5cm}, width=0.45\textwidth]{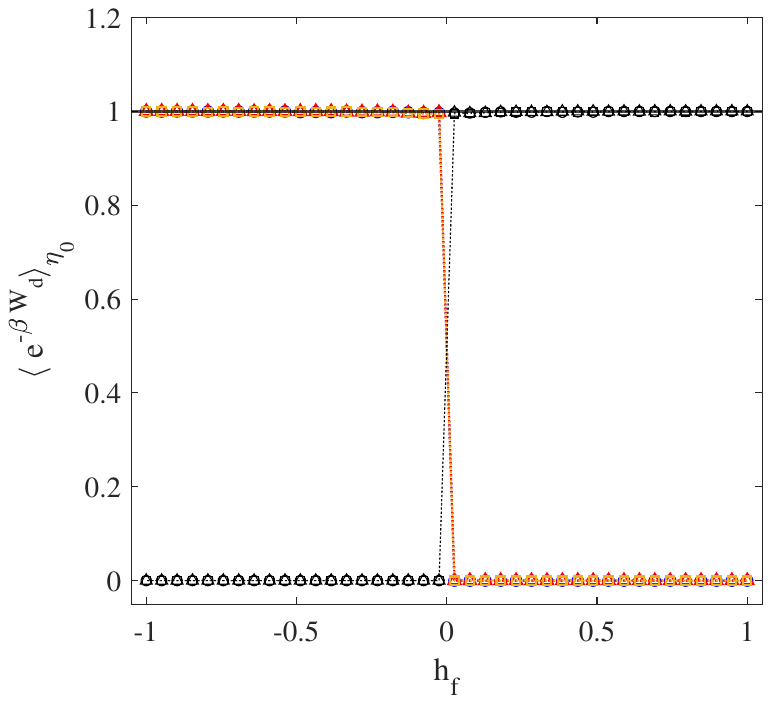}} \hspace{0.4cm}
        \subfigure{\includegraphics[trim = {3cm 9.5cm 3cm 9.5cm}, width=0.45\textwidth]{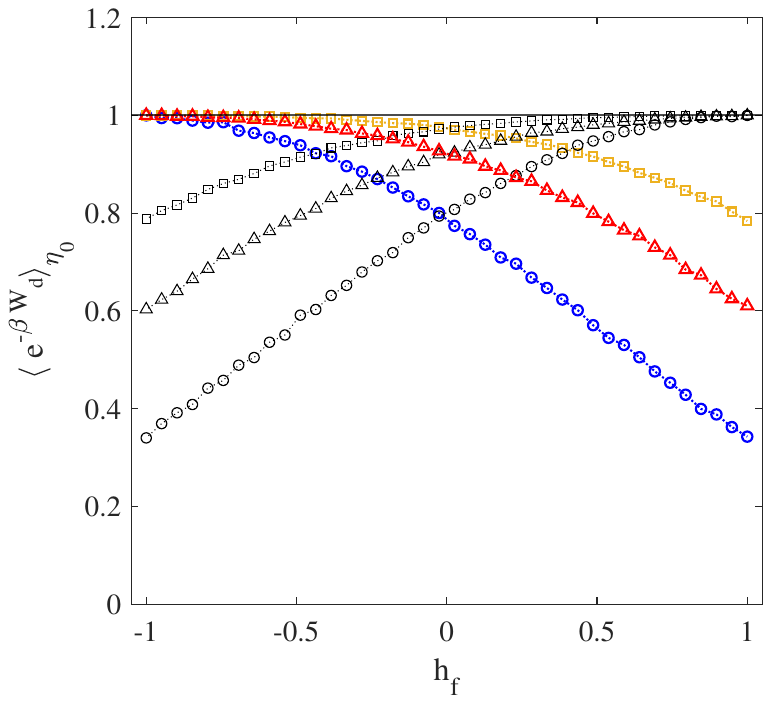}} 
\caption{$\langle e^{-\beta W_d} \rangle_{\eta_0}$ as a function of $h_f$,  Eq.\  \eqref{eq:final2}, from Monte Carlo simulations of a 2D Ising model on a square grid of linear size $L=10$. Here, $J=1$, hence $\beta_c\simeq 0.4406$, and periodic boundary conditions are applied on both horizontal and  vertical directions.
Simulations have been performed with $h_i=-1$ (coloured symbols) and $h_i=1$ (black symbols), for $\beta=2.5>\beta_c$ (left panel) and $\beta=0.1<\beta_c$ (right panel).
Circles, triangles and squares correspond to $\bar{\mu}=(\max\lbrace\mu_0\rbrace+\max\lbrace\mu_1\rbrace)/(2 k)$
with $k=5, 25, 100$, respectively.  
The horizontal black line at $1$ 
represents in both panels the case with $\bar{\mu}=0$.}  
        \label{fig:avexpWd_Ising}
    \end{figure}
Figure \ref{fig:avexpWd_Ising} shows that $\langle e^{-\beta W_d}\rangle=1$ when $\bar{\mu}=0$, for both $\beta<\beta_c$ and $\beta>\beta_c$, cf.\ Eq.\ \eqref{eq:avexpWd_ninS}. Therefore, the phase transition in the Ising model is not detected by the JE in one-step transformations. For the reduced the state spaces, corresponding to cutoffs $\bar{\mu}>0$, the situation changes as evidenced by 
the left panel of Fig.\ \ref{fig:avexpWd_Ising}. The phase transition is properly captured by $\langle e^{-\beta W_d}\rangle_{\eta_0}$, which drops abruptly from the value $1$ to a value close to $0$ when crossing $h_1=0$, for $h_0=-1$ and with $h_1 \in [-1,1]$. An analogous behavior is also observed when $h_0=1$ and $h_1$ is varied in the same interval. It is worth recalling that, in the Curie-Weiss approximation of the Ising model, the value of the magnetization density $m$ is obtained from the solution of the mean field equation $m=\tanh{[\beta (m+h)]}$, which, for $\beta>\beta_c$, gives rise to metastable branches \cite{Presutti}. One such branch corresponds to positive values of $m$ in the interval $[-h_c,0]$, and another branch corresponds, symmetrically, to negative values of $m$ in the interval $[0,h_c]$, with $h_c>0$. This is not visible in the left panel of Fig. \ref{fig:avexpWd_Ising}, which  shows that for $h_0=\pm 1$ the discontinuity occurs in both cases sharply at $h_1=0$.
This is
a consequence of our choice of 
initial datum for the Monte Carlo simulations, that mitigates the effects of the two metastable branches. More precisely, as initial datum we took $\sigma_j=-1$ for $h_1<0$ and $\sigma_j=1$ for $h_1>0$, for all $j\in \Lambda$,  
producing a distribution $P(m)$ sharply concentrated near $m=-1$ for $h_1<0$ and near $m=1$ for $h_1>0$. The discontinuity at $h_1=0$ therefore stems from Eq.\ \eqref{eq:bound}, which rules the behavior of $\langle e^{-\beta W_d} \rangle_{\eta_0}$ for $\beta>\beta_c$, where one can assume $\mathcal{I}=\emptyset$.

In the left panel it is also interesting to note that, for $\beta=2.5>\beta_c$, the different sets of data points corresponding to different values of $\bar{\mu}$ fully overlap with one another and display the same discontinuity at $h_1=0$.

For $\beta<\beta_c$ the phase transition in the Ising model is 
instead absent, and the magnetization density $m$ solving the mean field equation becomes a continuous function of $h$, that vanishes at $h=0$.
As a result, the data points shown in the right panel of Fig. \ref{fig:avexpWd_Ising} show that $\langle e^{-\beta W_d} \rangle_{\eta_0}$ no longer presents a sharp discontinuity at $h_1=0$, as this quantity decreases in a continuous fashion while increasing the difference $|h_0-h_1|$. The effect of varying the value of $\bar{\mu}$ is clearly visible in the right panel. Again, the JE is not affected by the variation of $h_1$.

\section{Conclusions}
\par\noindent
\label{s:Conclusions}

In this study, we propose one way of benefiting from incomplete statistics or finite size effects, that result in a reduction of the 
state space. As this is often an
unavoidable fact, we show that it can provide valuable insight, if associated with exponential observables. In particular, we have considered the Jarzynski Equality and its instantaneous transformation variant (also known as Free Energy Perturbation), to identify the critical values of the parameters corresponding to a first order phase transition. Smooth variations of states have also been characterized. This is due to the sensitivity of exponentials of extensive quantities, when averaged over the available subsets of the system state space, which is usually considered a hindrance.

Interestingly enough, our study show that reducing the amount of available information about the system can enhance our understanding of its critical behaviors. In fact, this is not obtained from a perfect sampling of states. The JE, in particular, 
results insensitive to parameters variations, hiding interesting features of the system at hand.

We illustrated our theoretical findings numerically analyzing two stochastic models: a modified Ehrenfest urn model and a classical 2D Ising model subject to an external magnetic field. 
In both cases, we observed a jump discontinuity in the quantity $\langle e^{-\beta W_d}\rangle_{\eta_0}$ occurring at the same value of the control parameter distinguishing different macroscopic phases in the thermodynamic limit of the model, remaining quite far from that limit. 
Similarly, when the phase transition is absent, as in the case of the Ising model at supercritical temperatures, the variation of state with the control parameter, is still revealed by a continuous variation of $\langle e^{-\beta W_d}\rangle_{\eta_0}$. 

Studying the behaviour of powers of $W_d$ we also found that our approach works even with 
variables different from those of fluctuation theorems. Indeed, it provides a new recipe to identify order parameters in the study of nonequilibrium phase transitions, that is particularly effective with truncated distributions.

Future works are intended to develop this approach, so that incomplete information on a given system may be used beyond the  problem of discovering phase transitions.

\vskip 10pt
\noindent{\bf Acknowledgements}
LR gratefully acknowledges support from the Italian Ministry of University and Research (MUR) through the grant PRIN2022-PNRR project (No. P2022Z7ZAJ) “A Unitary Mathematical Framework for Modelling Muscular Dystrophies” (CUP: E53D23018070001). 
This work has been performed under the auspices of Italian National Group of Mathematical Physics (GNFM) of the Istituto Nazionale di Alta Matematica (INdAM).

\bibliographystyle{ieeetr}
\bibliography{biblio}

\end{document}